# The optimal form of the scanning near-field optical microscopy probe

**N M Arslanov**
Zavoisky Physical-Technical Institute of the Russian Academy of Sciences,
Sibirsky Trakt str. 10/7, Kazan, 420029, Russia

**E-mail:** narslan@mail.ru

**Abstract.** A theoretical approach to determine the optimal form of the near-field optical microscope probe is proposed. An analytical expression of the optimal probe form with subwavelength aperture has been obtained. The advantages of the probe with the optimal form are illustrated using numerical calculations. The conducted calculations show 10 times greater light throughput and the reception possibility of the more compactly localized light at the output probe aperture which could indicate better spatial resolution of the optical images in near-field optical technique using optimal probe.

**Keywords:** scanning near-field optical microscopy, optimal form of probe, irregular waveguides and fibers.

**PACS numbers:** 07.79.Fc; 87.64.Xx; 42.25.Bs, 42.25.Gy

## 1. Introduction

The rapid development of modern technologies has drawn the researchers' attention to the physics of the surface phenomena [1]. Due to possibility of the reception ultrahigh resolution of the surface images the near-field optical microscopy technique has gained a great importance. In this technique light falls into metal coated taper dielectric fiber and goes through aperture with lateral size much less than light wavelength. Due to this light could be localized within a small sample area and one can get the optical images of surfaces with ultrahigh resolution. A non-propagating (evanescent) field that plays the main role in interaction with the surface [2] exists in a nearby zone of the output probe aperture (figure 1). The interaction of such highly localized in space polarized field with surface and individual molecule became the object of the intensive study [3, 4]. Characteristics of the probe have considerable importance in achievement of the efficient light localization [5]. Choice of the optimal form would allow one to reduce the size of the light spot in maximum magnitude and save the light intensity. Thus on the initial stage it is necessary to understand the interference picture of the light field in the output probe cross-section as it depends on forms of the probe used. It should also be noted that optical probes with small size of the intensive light spot would open new areas of the investigations such as nanometer nonlinear optics.

At present there are only experimental empirical ways of the probe optimization [see for example ref.1 and therein]. One of these schemes of the probe improvement is the reception smooth probe wall technology realized by means of polymeric covering dielectric fiber [6]. To increase the light throughput, probe aperture is processed by ion beam, avoiding absorbed impurity of the aluminum coating [7]. It is also offered to use the triangular form of the output aperture [8] and reception of coaxial probe [9] for increasing the resolution of images.

Several main trends draw attention in theoretical investigations [10-16], however, due to difficulties in the calculation of electromagnetic field in complex spatial geometry in existing theoretical works a question about the optimal form determination of the probe and influence of its form on parameters of output light, its polarization and intensity has not been set [1]. The short review of these studies is also presented in papers [17-18]. Searching for the optimal form of the probe with accounting for real physical parameter is the subject of the presented paper. B. Z. Katzenellenbaum theory of the cross-sections [19] has been used in the investigations of the light behaviour in a narrowing probe. This method is the most suitable for calculations since it enables the possibility to conduct numerical modeling and get comprehensible physical picture of the light spreading in the real probe. The proposed theoretical approach has allowed us to set the problem of finding the optimal form of the near-field optical microscopy probe, which provides the maximum intensity of output radiations under given size probe apertures. The conducted calculations have shown that probes with optimal form allow one to increase the light intensity approximately in 10 times with the same size of the output probe aperture.

## 2. Analytical expression of optimal probe profile

The theory of the cross-sections for non-uniform waveguides [19] has been used at the statement of the problem on determination of the optimal form of the probe. In this method, waveguide with the same radius as the probe has in this section is compared with each cross-section narrowing probe. The light field is introduced as set of the modes:

$$\vec{E}(z) = \sum_{-\infty}^{\infty} P_j(z)\vec{E}^j(z), \quad \vec{H}(z) = \sum_{-\infty}^{\infty} P_j(z)\vec{H}^j(z) \tag{1}$$

with coefficients of the decomposition obeying to the system of equations:

$$\frac{dP_j(z)}{dz} - ih_j(z)P_j(z) = \sum_{v=-\infty}^{v=\infty} S_{jm}(z)P_m(z), \tag{2}$$

where $S_{jm}$ – is a couple coefficients between the modes:

$$S_{jm}(z) = \frac{a(z)}{2h_j(z)(h_j(z) - h_m(z))} \frac{da(z)}{dz} \oint_C d\varphi \left(1 - \frac{\varepsilon_0}{\varepsilon}\right)(H_z^j H_z^m - H_\varphi^j H_\varphi^m + \varepsilon_0 E_r^j E_r^m). \tag{3}$$

It has been taken into account here, that near-field optical microscopy probes can be considered as nonmagnetic $\mu = \mu_0 = 1$, and the probe is uniform narrowing on cross-section. r, ϕ, z – is a cylindrical coordinates, a(z) - is a waveguide radius, $da(z)/dz$ tangent inclination angle of the probe wall, $h_j(z)$ - is a wave number of the mode j. For waveguide with walls from ideal metal:

$$h_j(z) = \left(k_0^2 \varepsilon_0 - v_j^2 / a(z)^2\right)^{1/2}, \tag{4}$$

where $k_0 = \omega/c$ - a wave number of the light, $\varepsilon_0$, $\varepsilon$ - dielectrical permeability of the probe core and metallic coating. In our numerical calculation $\varepsilon_0 = 2.16$. $v_j$ - j-th root of Bessel functions for TM modes, $\mu_j$ - j-th root of derived Bessel function for TE modes.

It is possible to divide near-field optical microscopy probe into 3 areas depending on mode wave number. The modes preferentially have spreading nature $\operatorname{Re} h_m > 0$ in the first area $k_0^2 a^2 \varepsilon_0 \gg v_m^2$ before critical cross-section $\tilde{a}(\tilde{z}_m)$. The critical cross-section $\tilde{a}(\tilde{z}_m)$ approaches in the second area $k_0^2 a^2 \varepsilon_0 \approx v_m^2$ for mode $h_m(\tilde{z}) \approx 0$. In the third area $k_0^2 a^2 \varepsilon_0 \ll v_m^2$ mode gains the evanescent nature ($\operatorname{Im} h_m \neq 0$).

Usually metallic probe coating is aluminum with $\varepsilon = -34.5 + 8i$ and has a thickness more, than skin layer with $\delta \approx 5-6$ nm thickness [20]. In probe due to finite wall conductivity $\zeta = \sqrt{\mu/\varepsilon}$ modes of the field (1) are expressed [18] through well known functions of the waveguide modes with ideal wall [20], but wave number TM modes is the form of [18]:

$$h_j^{TM}(z) = \left(k_0^2 \varepsilon_0 - \frac{v_j^2}{a^2(z)} + \zeta \frac{2ik_0 \varepsilon_0}{a(z)}\right)^{1/2}, \tag{5}$$

and for TE modes:

$$h_j^{TE}(z) = \left(k_0^2 \varepsilon_0 - \frac{\mu_j^2}{a(z)^2} + \frac{2i\zeta}{k_0(\mu_j^2 - n^2)} \frac{\mu_j^2}{a(z)^3} + \zeta \frac{2ik_0 \varepsilon_0}{(\mu_j^2 - n^2)} \frac{n^2}{a(z)}\right)^{1/2}. \tag{6}$$

Therefore, there are a greater number of the modes, than in probe with ideal wall [22] even in the third area $k_0^2 a^2 \varepsilon_0 \ll v_m^2$.

The system (2) couples modes spreading from broad part of probe and reflected modes because of the narrowing of the wall of the probe. The boundary condition for the total field (1) is satisfied in close proximity to the tilted wall [19]. Because of rapid change of the boundary conditions in the narrowing probe, the modes existence condition and regimes of their interaction greatly change, influencing the physical picture of the light field spreading. The light modes of the field are reflected, interact with mode of HE or EH types in the same and opposite directions when spreading in the probe. This leads to changing of the modes spatial structure and their amplitudes ratio [18] on output probe aperture that defines the structure and output light intensity and resolution of the near-field optical microscopy technique.

The interaction between modes noticeably increases in the narrowing probe. That greatly changes distribution of energy in the probe on modes of the same direction and leads to modes decay [18]. From couple coefficients ratio it is possible to draw a conclusion that there are areas in near-field microscopy probe, where the main role in the carrying dynamic of the radiation is played by either interaction nearby modes, since couple of the main mode with nearby mode has the greatest importance $S_{12} \gg S_{1,n>2}(z), S_{1,-i}(z)$, or reflection of the mode $S_{1,-1} \gg S_{1,n>1}(z)$ [18, 19]. The main mode has the least decay when spreading. So the general passing of the modes begins to decrease rapidly after the achievement of the critical cross-section $\tilde{z}_m$ (when is executed $h_m(\tilde{z}_m) \approx 0$) in corresponding ideal probe by main mode. The idea of the proposed approach for searching of the optimal probe geometry is as follows. Since high modes have a greater decay in probe then at excitement of the main mode in probe its optimal form must provide the minimum dissipation of this mode in modes of the higher order, minimum absorption in wall and reflection. The optimization of these mechanisms makes it possible to get the optimal probe profile.

We shall find the optimal profile in the probe area, when the nearby mode exists. The system (2) is possible to be reduced to the form:

$$\frac{d}{dz}\tilde{P}_1(z) = \tilde{S}_{1,m}(z) \cdot \tilde{P}_m(z), \tag{7}$$

where under m=2 minimize the interaction with nearby mode, and under m=-1 minimize the reflection.

$$P_j(z) = \tilde{P}_j(z) \exp\{i\int_0^z h_j(z')dz'\}, \tag{8}$$

$$\tilde{S}_{1,m}(z) = S_{1,m} \exp\{i\int_0^z \{h_m(z') - h_1(z')\}dz'\}. \tag{9}$$

It is possible to neglect remaining terms since either $\tilde{S}_{12} \gg \tilde{S}_{1n>2}(z), \tilde{S}_{1,-i}(z)$, or $S_{1,-1} \gg S_{1,n>1}(z)$. Equation for probe radius a(z) depending on longitudinal coordinate z is possible to define, imposing condition:

$$\delta \int_0^L dz |\tilde{S}_{1,m}(z,a,a')|^2 \bigg|_{\substack{a(0)=a1 \\ a(L)=a2}} = 0, \tag{10}$$

which points to minimum interaction of the main mode $|\tilde{S}_{1,m}(z)|$ over all narrowing part of the probe. In this case the energy of the electromagnetic field, spreading to the probe output, will remain only in the main (first) mode.

For TM modes equation (10) is reduced to equation of the type (refer to appendix A):

$$\frac{d^2a(z)}{dz^2} - \frac{1}{a}(\frac{da(z)}{dz})^2(1 - \frac{A_{1,m}(z,\lambda)}{2a^2}) - \frac{da(z)}{dz} \text{Im}\, h_{m,1}(z) = 0, \tag{11}$$

where $h_{m,1}(z) = h_m(z) - h_1(z)$ and:

$$A_{1,2}(z,\lambda) = \left(\frac{\zeta i k_0 a \varepsilon_0 (h_2^2 + h_1^2) - v_2^2 h_1^2 - v_1^2 h_2^2}{h_1 h_2 (k_0^2 \varepsilon_0 - h_1 h_2)} + \frac{v_1^2 - \zeta i k_0 a \varepsilon_0}{h_1^2 h_2}(h_1 - h_2) - \frac{(v_1^2 - v_2^2)}{h_2(h_1 - h_2)}\right) + h.c. \tag{12}$$

$$A_{1,-1}(z,\lambda) = -\frac{k_0^2 \varepsilon_0}{h_j^2} \frac{2(v_{nj}^2 - \zeta i k_0 a \varepsilon_0)}{(k_0^2 \varepsilon_0 + h_j^2)} + h.c. \tag{13}$$

When wall impedance of the probe can be considered negligibly small $\zeta \to 0$ at small transverse size of the probe $k_0^2 a^2 \varepsilon_0 \ll v_2^2$ or at large transverse size of the probe $k_0^2 a^2 \varepsilon_0 \gg v_2^2$ so that in absence of the critical radius $\tilde{a}(\tilde{z})$ of the second mode (in critical cross-sections $\tilde{z}$ is executed $h_m(\tilde{z}_m) = 0$), the equation (11) has an analytical solution. In this case $\frac{A_{1,m}(z,\lambda)}{2a^2} \ll 1$ and equation (11) is reduced to the form:

$$\frac{d^2a(z)}{dz^2} - \frac{1}{a(z)}\left(\frac{da(z)}{dz}\right)^2 - \frac{K}{a(z)}\frac{da(z)}{dz} = 0, \tag{14}$$

for $k_0^2 a^2 \varepsilon_0 \ll v_j^2$, where:

$$K = \begin{cases} v_2 - v_1 & TM \\ \mu_2 - \mu_1 & TE \end{cases}. \tag{15}$$

The last term in (14) describes decay in the walls of the probe. In the case of $k_0^2 a^2 \varepsilon_0 \gg v_2^2$:

$$\frac{d^2a(z)}{dz^2} - \frac{1}{a(z)}\left(\frac{da(z)}{dz}\right)^2 = 0. \tag{16}$$

The analytical form of the probe, obtained by solution of the equations (14) and (16) for TM and TE modes in probe with the ideal walls has the form:

$$a(z) = \frac{\lambda}{C_1}(K\frac{L}{\lambda} + C_2 e^{C_1 z/L}), \quad (17)$$

$$a(z) = a_1 e^{C_3 z/L}. \quad (18)$$

In close proximity of the modes critical radiuses $\widetilde{a}$, when is executed $h_1(\widetilde{z}_1) \approx 0$ or $h_2(\widetilde{z}_2) \approx 0$, term $\frac{A_{1,m}(z,\lambda)}{2a^2}$ becomes comparable to or even greater than unit. For the area of the critical cross-section $\widetilde{z}_1$ of the first mode is performing $A(z,\lambda) \approx -\frac{v_j^2}{h_j^2} + h.c. \approx -\frac{v_j^2 a}{|\zeta|^2 \sqrt{|\varepsilon'|}k_0 \varepsilon_0}$ and for the area of the critical cross-section of the second mode $\widetilde{z}_2$ is performing $A(z,\lambda) \approx \frac{v_j^2 v_m^2}{k_0^2 \varepsilon_0 a^2 h_j}(\frac{1}{h_m} + h.c.) \approx \frac{2v_j^2 v_m^2 \sqrt[4]{|\varepsilon'|}}{k_0^2 \varepsilon_0 \sqrt{v_m^2 - v_j^2}\sqrt{2k_0 a \varepsilon_0}}$. Due to finite wall conductivity the equation (11) has continuous solutions as decompositions in series of a small parameter for each of neighborhoods $\widetilde{z}_1$ and $\widetilde{z}_2$ that are joined on the boundary of these areas. Since areas of the critical cross-sections in probe are very small, then corresponding solution in the neighborhood $\widetilde{z}_1$ can be replaced by approximation (17) and solution in the neighborhood $\widetilde{z}_2$ can be replaced by the solution of (17) and (18) joining in section $\widetilde{z}_2$.

Thereby, the general analytical solution (17), (18) for the probe, narrowing from the initial radius $a_1$=500 nm to output $a_2$=50 nm at probe length L=450 nm for light with wavelength $\lambda$ = 500 nm, has the form:

$$a(z) = a_1 e^{C_3 z/L} \theta(\widetilde{z} - z) + \frac{\lambda}{C_1}(K\frac{L}{\lambda} + C_2 e^{C_1 z/L})\theta(\widetilde{z} - z), \quad (19)$$

and had been joined in critical cross-section for the second mode $\widetilde{z}_2$ with radius $\widetilde{a}_2(\widetilde{z} = 244nm) = \frac{v_2}{k_0 \sqrt{\varepsilon_0}}$=298.8 nm, where $\widetilde{Z}$ =244 nm had been defined from the numerical solution (11), and constant coefficients are defined from boundary conditions $a(0) = a_1$, $a(\widetilde{z}) = \widetilde{a}$, and $a(L) = a_2$: $C_1 \approx 4.066, C_2 \approx -0.041, C_3 = \frac{L}{\widetilde{z}} \ell n \frac{\widetilde{a}}{a_1} \approx -0.948$,

$\theta(z) = \begin{cases} 1, z \geq 0 \\ 0, z < 0 \end{cases}$.

In figure 2 the exact numerical solution (11) and approximate analytical solution (19) are presented. The difference between the numerical solution (11) and analytical solution (19) is around 1-2% and reaches 5% (~2.5 nm) in small area near output aperture. The comparison of the solutions demonstrates fairness of the assumed approximations for the subwavelength probe parameters at the construction of the analytical solution (19). Thereby, the analytical expression (19) is a good approximation of the equation (11) solution and describes the optimal form a subwavelength probe.

At the small radiuses area modes wavenumbers (4-6) has significant imaginary part and large couple coefficients (3) between modes that leads to essential energy dissipation of the main mode and absorption in walls [5, 18, 19, 21]. Thence follows the important conclusion that optimization in the end subwavelength area of the probe will lead to the essential increase of the light throughput and the improvement its spatial features. Using the obtained analytical expression (19), the numerical solution of the probe optimal form and other typical probe forms, the spreading of TM light field in small end area of the probe is investigated and the comparisons of output light parameters are performed and the advantages of the optimal form of the probe (19) are illustrated.

In figure 3 the next forms of the probe are presented: rectilinear $a(z) = a_1 + z\frac{a_2 - a_1}{L}$, analytical expression (17) extended on the whole area of the probe at $C_1 \approx 2.618, C_2 \approx -0.185$, exponential probe narrowing (18) at $C_3 = \ell n \frac{a_2}{a_1} = -2.303$, and probe profile, which is used in works [14]:

$$a(z) = \left(1 + \exp[-\frac{C_8(z-L)}{L}]\right)^{1/2}\left(\frac{1}{a_\infty^2} + \frac{1}{a_1^2}\exp[-\frac{C_8(z-L)}{L}]\right)^{-1/2}, \quad (20)$$

where $a_\infty = \left(\dfrac{2}{a_2^2} - \dfrac{1}{a_1^2}\right)^{-1/2}$ =35.44 nm, $C_8$=10 and optimal form, obtained by numerical and analytical solution (19) of the equation (11) for light with λ = 500 nm. It must noted that in the area $k_0^2 a^2 \varepsilon_0 < v_j^2$ the optimal form of the subwavelength probe noticeably differs from rectilinear and exponential probe profile and depends on the light wavelength λ.

The energy flow carried by all field modes along the probe axis is:

$$\vec{S}(z,\lambda) = \dfrac{c}{8\pi} \text{Re} \iint_\Sigma [\vec{E}\vec{H}^*] d\vec{S}, \qquad (21)$$

where fields are defined by the expression (1). The intensity of the light field in the probe is:

$$I(z,\lambda) = \iint_\Sigma |\vec{E}|^2 \, dS. \qquad (22)$$

In figure 4 throughput of the TM field energy flow at light wavelength changing within the range from 300 nm to 1000 nm are presented. The field energy throughputs for the probe with the linear form were determined by experimental and approximate theoretical methods in [1, 6, 7] for some light wavelengths and are in agreement by the order of magnitude with our calculations, derived theoretically for the first time for the whole spectral range. From the obtained dependencies it is possible to see that the optimal probe gives the greater energy flow of the light field on the output aperture. The comparison with the linear form of the probe shows that throughput of the $TM_{0m}$ field power can be increased by 10 times.

**3. The spatial structure of the near-field in probe output with optimal form**
The probe form significantly changes the behaviour of an evanescent field [18]. At modes interferences the output intensity differs from simple addition of the intensities of all modes [18]. The theoretical description of the near field has also great importance to increase the resolution of the near-field optical microscopy technique. It is possible to control the distance between the probe and the sample by using modern methods of the control [1]. In case of small sufficiently distance the resolution ability of the spatial measurements of surface will be defined by the spatial structure of the light field in the output probe aperture, rather then by the diameter of the aperture.

In figures 5, 6 intensities structure of longitudinal $I_z(r, z=L)$ and transverse components $I_r(r, z=L)$ of TM field on output probe aperture with optimal and linear form of the narrowing for wavelength 500 nm of spreading field is presented. Such theoretical distributions of intensities on cross-section of the output probe aperture were obtained for the first time [18] and are in good accordance with experimental results of the angular distribution diagram measurement of the output light polarization in the far-field [22] at the linear polarized light falling into the probe with wavelength 633 nm. However, the theoretical method allows us to study the structure of the field precisely in the near zone.

It is seen in the figures that there are areas with prevalence of longitudinal $E_z$ or transverse $E_r$ field in cross-section of the probe aperture. However, longitudinal light field defines namely the structure of the near-field light. Depending on locations of the sample molecule relatively to cross-sections of the probe the characteristics of the interaction of the light with molecule will be changed. So, surface interactions with longitudinal light field will define the images resolution, gained by near-field optical microscopy technique. We note that the best resolution can provide the probe with high-angle $70^o < \alpha < 75^o$ walls on the output probe area [18] and with high sufficiently magnitude of output intensity. The probe with optimal form in its possibilities combines these characteristics since it has the best output power (see fig. 4) and enough angle corner $\alpha \approx 71^0$ that provides ultrahigh interference compression of longitudinal and transverse component of the light field. The comparison of the width intensities on figures 5, 6 shows that using the probe with optimal form can increase the resolution in 3 times in comparison with linear narrowing and will allow to reach the resolution in 20 nm at aperture diameter 100 nm.

**4. Conclusion**
In the present work the approach to determine optimal form of the near-field optical microscopy probe is proposed. The developed approach makes it possible to calculate the light field at different geometric and real physical probe parameters. The analytical expression of the optimal form of the probe with subwavelength output aperture is obtained.

The comparison of the throughput and spatial structure of the light field in near zone at output probe aperture for typical forms of narrowing has been made numerically to illustrate the advantages of the optimal probe. This research has shown that in comparison with linear probe form using the optimal probe allows one to increase the intensity of the field at the output probe aperture by 10 times and carry out the experiments with such intensive laser radiation avoiding destroying the probe. We shall note that it is possible to use the laser radiation 90% smaller at maintenance intensities of the light at the output probe aperture with optimal form.

Herewith, the optimal form of the probe provides greater locality and magnitude of longitudinal and transverse light field in distribution at output cross-section of the probe aperture. At diameter of the aperture 100 nm intensity of $TM_{0m}$ light field in cross-section of the probe is localized in the area of 15-20 nm. This must allow one, in particular, to raise greatly the spatial resolution of the near-field optical microscopy technique, especially at the development of the appropriate methods of deconvolution.

**Acknowledgments**

The author is grateful to S. A. Moiseev for discussion and help and to A. A. Kalachev for deep interest to work and useful advice. This work was supported by the Russian Foundation for Basic Research grants № 03-03-96214, № 00-15-97410, Tatarstan NIOKR № 06-6.3-343/2005.

**Appendix A**

We find the equation on radius for TM field. The couple coefficient (9) has the form:

$$\tilde{S} = \frac{(k_0^2 \varepsilon_0 - h_j h_m)}{h_j(h_j - h_m)} \frac{da(z)}{a(z)dz} \exp(i\int_0^z dz(h_m - h_j)). \tag{A.1}$$

For functional variation (10) we obtain the Euler equation:

$$G(z)\frac{dJ(z,a,a')}{da} - \frac{d}{dz}(G(z)\frac{dJ(z,a,a')}{da'}) = 0, \tag{A.2}$$

where:

$$G(z) = |\exp\{i\int_0^z \{h_2(z') - h_1(z')\}dz'\}|^2, \tag{A.3}$$

$$J(z,a,a') = |S(z,a,\frac{da}{dz})|^2 = \frac{B_{1,m}(a(z))}{a^2(z)}\left(\frac{da(z)}{dz}\right)^2, \tag{A.4}$$

$$B_{1,m} = \frac{(k_0^2 \varepsilon_0 - h_j h_m)}{h_j(h_j - h_m)}\left(\frac{(k_0^2 \varepsilon_0 - h_j h_m)}{h_j(h_j - h_m)}\right)^*, \tag{A.5}$$

$$h_{-m}(z) = -h_m(z). \tag{A.6}$$

Whence, using:

$$\frac{1}{B_{1,m}(z,\lambda)} \frac{dB_{1,m}(z,\lambda)}{da} = \frac{A_{1,m}(z,\lambda)}{a(z)^3}, \tag{A.7}$$

where

$$A_{1,2}(z,\lambda) = \left(\frac{\zeta i k_0 a\varepsilon_0(h_m^2 + h_j^2) - \nu_m^2 h_j^2 - \nu_j^2 h_m^2}{h_j h_m(k_0^2\varepsilon_0 - h_j h_m)} + \frac{\nu_j^2 - \zeta i k_0 a\varepsilon_0}{h_j^2 h_m}(h_j - h_m) - \frac{(\nu_j^2 - \nu_m^2)}{h_m(h_j - h_m)}\right) + h.c. \tag{A.8}$$

$$A_{1,-1}(z,\lambda) = -\frac{k_0^2\varepsilon_0}{h_j^2}\frac{2(\nu_{nj}^2 - \zeta i k_0 a\varepsilon_0)}{(k_0^2\varepsilon_0 + h_j^2)} + h.c., \tag{A.9}$$

we obtain the equation on radius (11).

In the area where $\frac{k_0^2\varepsilon_0 a^2}{\nu^2} \ll 1$ and $\frac{k_0^2\varepsilon_0 a^2}{\nu^2} \gg 1$ we have $\frac{A_{1,m}(z,\lambda)}{2a^2} \ll 1$. In the area $\frac{k_0^2\varepsilon_0 a^2}{\nu^2} \ll 1$ the energy dissipation to nearby mode dominates and equation (11) will have the form (14). In the area $\frac{k_0^2\varepsilon_0 a^2}{\nu^2} \gg 1$ the equation (11) for both values m=2,-1 is reduced to equation (16).

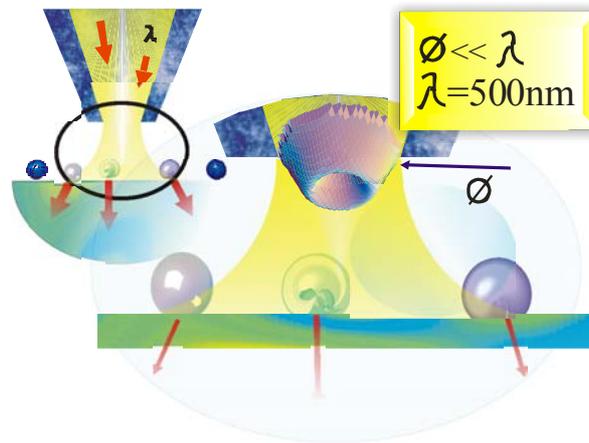

Figure 1. The near-field structure at the output aperture of near-field optical microscopy probe.

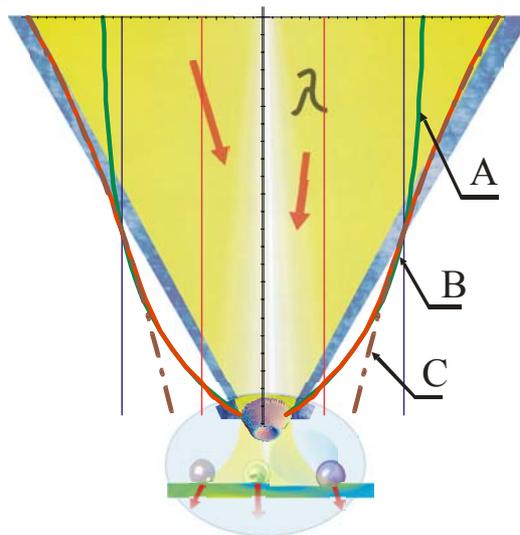

Figure 2. The comparison of the analytical solution (19) and the numerical (11). The straight lines show the critical radiuses of the first and the second mode.

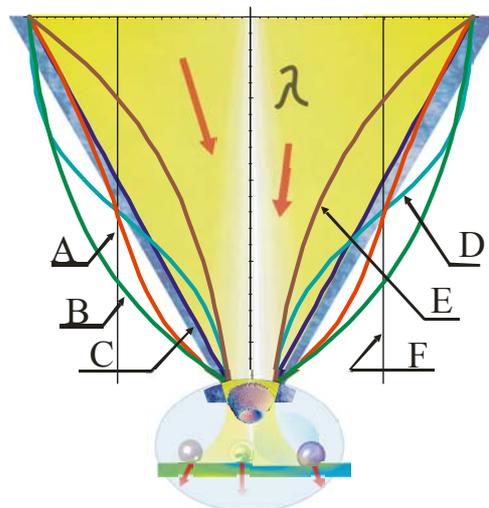

Figure 3. Forms of the probe for TM modes: A - optimal for $TM_{0m}$ (19), C - linear, D-profile (20), E - exponential (18), and B - the expression (17) extended on the whole area of the probe. The horizontal curve F shows the critical radius of the second mode, where solutions (17) and (18) are joined.

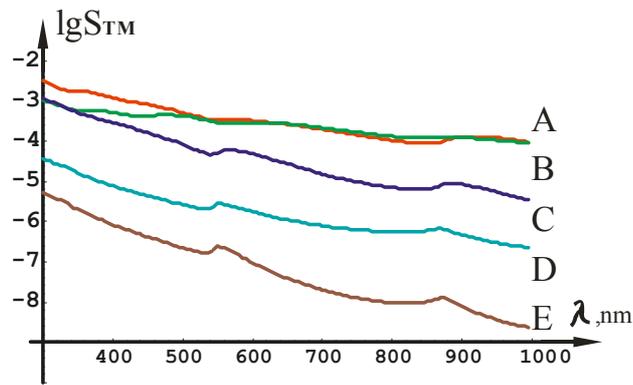

Figure 4. Total power throughput of the three modes related to the first mode powers at the initial probe aperture. For chosen probe parameters interaction between magnetic and electric HE and EH type modes has been neglected [18]. At the probe input the exciting field consisted of the first mode: $TM_{01}$. The forms of the probe are marked: A - an optimal form, B - analytical, C - linear; D - a form described by (20), E - exponential.

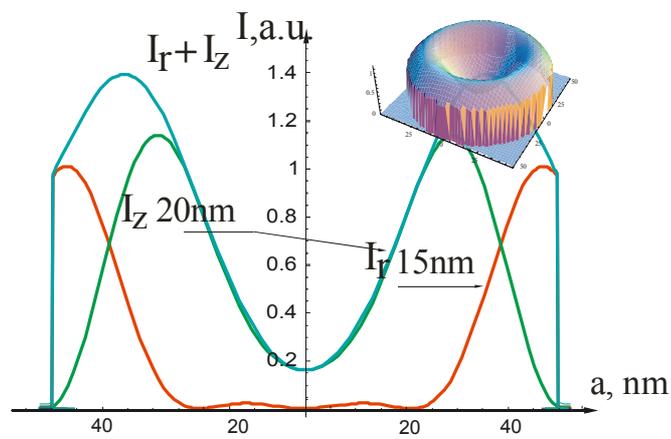

Figure 5. The intensity of longitudinal $I_z(r)$ and transverse $I_r(r)$ component of $TM_{0m}$ field on output aperture of the probe with optimal form for $\lambda = 500$ nm. (The arrows show half width of intensities on half height).

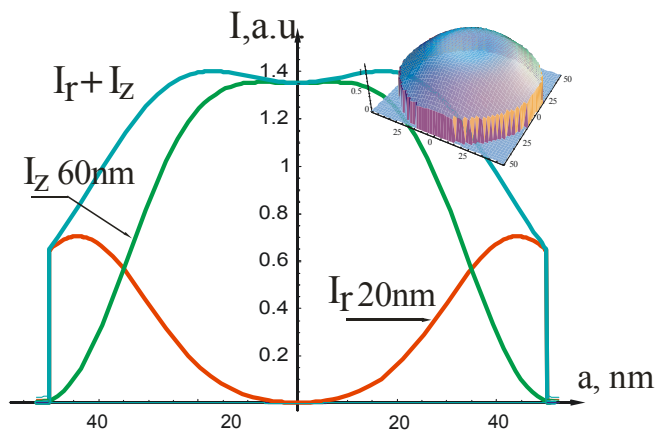

Figure 6. The intensity of longitudinal $I_z(r)$ and transverse $I_r(r)$ component of $TM_{0m}$ field on output aperture of the probe with linear form for $\lambda = 500$ nm. (The arrows show half width of intensities on half height).